# Infection fatality ratio of SARS-CoV-2 in Italy

Piero Poletti[1,*], Marcello Tirani[2,3,*], Danilo Cereda[2], Filippo Trentini[1], Giorgio Guzzetta[1], Valentina Marziano[1], Sabrina Buoro[2,4], Simona Riboli[5], Lucia Crottogini[2], Raffaella Piccarreta[6], Alessandra Piatti[2], Giacomo Grasselli[7], Alessia Melegaro[6], Maria Gramegna[2], Marco Ajelli[8,#], Stefano Merler[1,#]

[*]corresponding authors: poletti@fbk.eu, marcello_tirani@ats-pavia.it
[#]senior authors

[1] Bruno Kessler Foundation, Trento, Italy
[2] Directorate General for Health, Lombardy Region, Milan, Italy
[3] Health Protection Agency of Pavia, Pavia, Italy
[4] Quality Management Unit, Papa Giovanni XXIII Hospital, Bergamo, Italy
[5] Department of Public Health, Experimental and Forensic Medicine, University of Pavia, Pavia, Italy
[6] Bocconi University, Dondena Centre for Research on Social Dynamics and Public Policy, Milan, Italy
[7] Dipartimento di Anestesia, Rianimazione ed Emergenza-Urgenza, Fondazione IRCCS Ca' Granda Ospedale Maggiore Policlinico, Milan, Italy
[8] Department of Epidemiology and Biostatistics, Indiana University School of Public Health, Bloomington, IN, USA

## Abstract

We analyzed 5,484 close contacts of COVID-19 cases from Italy, all of them tested for SARS-CoV-2 infection. We found an infection fatality ratio of 2.2% (95%CI 1.69-2.81%) and identified male sex, age >70 years, cardiovascular comorbidities, and infection early in the epidemics as risk factors for death.

## Main text

The proportion of infections resulting in a fatal outcome, known as infection-fatality ratio (IFR), and the associated risk factors are still poorly quantified for the newly emerged SARS-CoV-2 [1,2]. Estimates of the IFR are key to evaluate the health impact of epidemics and the effectiveness of control strategies [3,4]. Estimates of the IFR conditioned on age, sex, and comorbidities are useful to project the expected number of deaths in populations characterized by different demographics and prevalence of chronic diseases, and to identify the most vulnerable segments of the population.

Given the high proportion of asymptomatic and pauci-symptomatic SARS-CoV-2 infections [5,6], it is difficult to estimate the IFR from surveillance data [2-4]. In addition, the proportion of ascertained cases over all infections can change dramatically across regions and over time [3,4]. The increasing availability of serological data can assist precise and direct measurements of the IFR. Literature estimates available so far are highly variable (ranging from 0.07% to 1.6%) [1-3,7,8], based on small non-random samples [8], data pooled from heterogeneous populations [1-3,7], or derived through modeling analyses [3].

In this study, we compute stratified IFR values for SARS-CoV-2 by combining different data on the same group of individuals. The study population comprises contacts of COVID-19 cases identified through contact tracing operation conducted in Lombardy, Italy between February and April 2020. For these subjects, laboratory results on reverse transcription polymerase chain reaction (RT-PCR) from nasopharyngeal swabs administered during the contact tracing activity are analyzed. This data is complemented with the results of an ongoing serological survey on the same group of individuals started on April 16, 2020. Finally, we also collected information on the presence of comorbidities (respiratory, cardiovascular, metabolic and oncological) and clinical outcomes of each case reported in the Lombardy linelist of COVID-19 patients (last update: June 8, 2020).

Data analyzed here represent a selection from a database of 62,881 contacts of COVID-19 cases. We selected only the contacts belonging to clusters (i.e., groups of contacts identified by one positive index case) where all individuals were tested against SARS-CoV-2 infection either through nasal swabs during the contact tracing operations or IgG serological testing retrospectively. In fact, to economize the saturated testing resources in Lombardy, from February 26, 2020 onward (i.e., shortly after the detection of the first locally-transmitted COVID-19 case in Italy) only symptomatic cotnacts of COVID-19 cases were tested via RT-PCR. On April 16, 2020, the regional health authorities started a serological survey aiming at completing the ascertainment of SARS-CoV-2 infections among all close contacts identified for each confirmed case. To avoid possible biases, we excluded index cases, who were often identified because of their symptoms and were thus possibly at higher risk of severe disease, from our analysis. The accuracy of IgG testing and RT-PCR used in our sample was assessed in [9,10]. The IFR was computed as the proportion of deaths occurred among all SARS-CoV-2 positive contacts identified in the considered sample, here defined as subjects with at least one laboratory confirmation of their infection.

Overall, we analyzed 5,484 contacts (median age 50, IQR 30-61; 43.7% males), of which 1,364 (25%) were tested only by an RT-PCR assay targeting different SARS-CoV-2 genes during the contact tracing activities [9]; 3,493 (64%) only by a serological assay for IgG neutralizing antibodies against S1/S2 antigens [10] at least a month after the reporting date of their index case; 627 (11%) were tested both with RT-PCR and serological assays. Out of 5,484 analyzed individuals, 2,824 resulted positive to SARS-CoV-2 (median age 53, IQR 34-64; 43.2% males), and among them 62 died with a COVID-19 diagnosis (median age 79, IQR 74-83; 53.2% males).

We found an IFR of 18.35% (95%CI 12.65-25.28%) for individuals aged 80 years or more, 3.62% (95%CI 2.45-5.13%) for those aged 60-79 years, and 0.16% (95%CI 0.03-0.48%) for those younger than 60 years (Table 1 and Figure S1 in Appendix). No deaths were recorded among individuals below 50 years of age. The overall IFR was estimated at 2.2% (95%CI 1.69-2.81%), but it decreased over time, ranging from 2.95% (95%CI 2.14-3.97%) for patients associated with clusters identified before March 16, 2020 (the median date of confirmation among index cases in the considered clusters) to 1.33% (95%CI 0.79-2.09%) for clusters identified afterwards. Such a difference was especially marked among infections over 80 years of age (IFR: 30.4% vs 8.1%). In our sample, 51 out of 62 deaths occurred in patients were affected by cardiovascular diseases. The IFR for infections with this comorbidity was at 16.14% (95%CI 12.26-20.67%). The overall IFR was 50% higher in males (2.70% vs 1.80%).

To identify the risk factors associated with fatal outcomes after SARS-CoV-2 infection we applied a generalized linear model (GLM with logit link) relating the observed outcome (death vs. survival) to the sex and the age group (0-59, 60-69, 70-79, 80+ years) of exposed individuals, to the presence of comorbidities (none, cardiovascular, others) and to the epidemic period associated with the observed outcomes (before or after March 16, 2020). We found that individuals younger than 70 years were at a significantly lower risk of death after infection compared to older patients (Tukey test: p-value<0.001); females were 55.2% less likely to die compared to males (95%CI 31.6-97.0%; Tukey test: p -value<0.001); the relative risk of death was 5.6 times higher for patients affected by cardiovascular comorbidities (95%CI 1.2-55.3) compared to otherwise healthy individuals, although this difference was not statistically significant (Tukey test: p -value=0.29). Finally, the risk of death was 62% lower (95%CI 31-80; Tukey test: p-value<0.001) during the second phase of the epidemic (Table S1 in Appendix).

Our estimate of the overall IFR (2.2%) is consistent with the range of estimates obtained for Spain (from 1% to 2% [2]), but higher than estimates reported in other studies [1-3,7,11]. Such differences may be due to the high proportion of older individuals in the Lombardy population (28.7% of people older than 60 years), or differences in the prevalence of chronic diseases. In particular, more than 80% of the deaths recorded in our sample occurred in patients with cardiovascular diseases, which represented the most common comorbidity among patients hospitalized in the early phase of the Italian epidemic [12]. Due to the limited sample size we are unable to provide solid estimated of the IFR stratified both by age and comorbidities. Finally, the high average IFR we found may also be related to the specific area and time period considered here, mirroring a health system under a severe strain because of the rapid, massive increase of patients requiring intensive care [12]. These conditions combined with the initial scant evidence on appropriate treatments of disease may well have affected the health system capacity to cope with severe cases. This hypothesis is supported by the remarkable difference in the IFRs estimated for clusters identified before and after March 16 (2.95% vs. 1.33%).

The age distribution of our sample reflects that of the Lombardy population (Figure S2 in Appendix). However, the data considered in this study represent a sample of individuals who were exposed to COVID-19 cases.

Therefore, the reported infection attack rates cannot be considered as representative of the immunity level caused by SARS-CoV-2 in this region.

A limitation of the presented analysis is that infections were identified in two different periods of time and using different tests (RT-PCR and IgG serological assays). In our sample, all contacts were followed for symptom onset, but while symptomatic individuals were tested by PCR shortly after their index case was identified, IgG serological tests were performed on individuals more than 1 month after their identification as contacts. These features minimize potential biases related to either the identification of infections and deaths or to seroconversion delays. However, out of 327 contacts tested by both RT-PCR and serology, 137 resulted negative to RT-PCR and positive to serology (Table S2 in Appendix). We explored to what extent failures in RT-PCR testing may impact estimates on the overall IFR, by considering a worst-case scenario where 41.9% (i.e. 137/327) of negative contacts who were tested with RT-PCR only (i.e., 732) are assumed to be positive. The estimated IFR in this case becomes 2.04% (95%CI 1.57-2.60%).

The estimates provided here can be considered a robust representation of the IFR during the COVID-19 epidemic in Lombardy - the most affected Italian region [9,12]. These results can be instrumental in evaluating the expected burden of possible future outbreaks. The indication on the key factors that are strongly influencing the SARS-CoV-2 IFR could be used to inform targeted interventions and possible future COVID-19 vaccination campaigns.

# Tables and Figures

|  | Contacts | SARS-CoV-2 positive | Deaths | Infection fatality ratio (%) |
|---|---|---|---|---|
| **Total** | 5,484 | 2,824 | 62 | 2.20 (95%CI 1.69-2.81) |
| Sex |  |  |  |  |
| Males | 2,398 | 1,220 | 33 | 2.70 (95%CI 1.87-3.78) |
| Females | 3,086 | 1,604 | 29 | 1.81 (95%CI 1.21-2.59) |
| **Age** |  |  |  |  |
| 0-19y | 692 | 304 | 0 | 0 (95%CI 0-1.21) |
| 20-49y | 1,951 | 885 | 0 | 0 (95%CI 0-0.42) |
| 50-59y | 1,241 | 648 | 3 | 0.46 (95%CI 0.1-1.35) |
| 60-69y | 867 | 494 | 7 | 1.42 (95%CI 0.57-2.9) |
| 70-79y | 485 | 335 | 23 | 6.87 (95%CI 4.4-10.12) |
| 80+ | 248 | 158 | 29 | 18.35 (95%CI 12.65-25.28) |
| **Comorbidity** |  |  |  |  |
| None | 122 | 113 | 1 | 0.88 (95%CI 0.02-4.83) |
| Cardiovascular # | 350 | 316 | 51 | 16.14 (95%CI 12.26-20.67) |
| Respiratory system # | 50 | 49 | 8 | 16.33 (95%CI 7.32-29.66) |
| Oncological # | 106 | 92 | 11 | 11.96 (95%CI 6.12-20.39) |
| Diabetes and other metabolic # | 93 | 79 | 13 | 16.46 (95%CI 9.06-26.49) |
| Unknown | 4,947 | 2,335 | 9 | 0.39 (95%CI 0.18-0.73) |
| **Epidemic period** |  |  |  |  |
| Before March 16, 2020 | 2,696 | 1,423 | 42 | 2.95 (95%CI 2.14-3.97) |
| After March 16, 2020 | 2,721 | 1,358 | 18 | 1.33 (95%CI 0.79-2.09) |
| Unknown | 67 | 43 | 2 | 4.65 (95%CI 0.57-15.81) |

# include patients with other comorbidities

**Table 1.** Sample description and IFR estimates by sex, age group, comorbidities, and epidemic period.

## Acknowledgments

PP, FT, GG, VM, and SM acknowledge funding from the European Commission H2020 project MOOD and from the VRT Foundation Trento project "Epidemiologia e trasmissione di COVID-19 in Trentino"

## Competing interests
The authors declare no competing interests.

## Contributions
PP,MA,SM conceived and designed the study. PP performed the analysis. MT,DC,SB,SR,LC,AP,MG collected data. PP,MT,DC,FT collated linked clinical–epidemiologic data. MT,DC verified all data. PP,GG,MA wrote the first draft. All authors contributed to data interpretation, critical revision of the manuscript and approved the final version of the manuscript

# Appendix

## Materials and Methods

### Data

The data analyzed in this study were taken from a database providing, among other details, information on sex, age, presence of comorbidities for close case contacts, the results of RT-PCR tests and serological tests (if any), and the clinical outcome of positive cases. In particular, the database was built by combining:
i) data records collected during the contact tracing activities conducted between February 21 and April 16, 2020 by the Lombardy healthcare agencies;
ii) results of RT PCR assays on nasopharyngeal swabs mainly administered to symptomatic contacts short after their identification during contact tracing;
iii) results from serological assays collected within a seroprevalence study started on April 16, 2020 and not yet completed, which mainly targets case contacts who were not tested by RT-PCR or resulted negative to RT-PCR;
iv) information (updated to June 8, 2020) on the clinical outcome of patients as available in the linelist of all COVID-19 laboratory confirmed cases in Lombardy.

Contact data collected after April 16, 2020 were excluded to avoid biases caused by delays in development of symptoms, reporting, and in seroconversion of infected individuals. The performed analysis is based on the serological test results obtained by May 25, 2020 and on the linelist of all COVID-19 cases detected in the region, as updated on June 8, 2020. In this study, we selected only contacts belonging to clusters with complete testing, i.e. clusters whose contacts all received at least one between a valid RT-PCR result or a valid serological result [1]. Clusters with case contacts with inconclusive serological results were excluded by the proposed analysis.

A close case contact was defined as either of the following:
- a person living in the same household as a COVID-19 confirmed case;
- a person having had face-to-face interaction with a COVID-19 confirmed case;
- a person who was in a closed environment (e.g. classroom, meeting room, hospital waiting room) with a COVID-19 confirmed case at a distance of less than 2 meters for more than 15 minutes;
- a healthcare worker or other person providing direct care for a COVID-19 confirmed case, or laboratory workers handling specimens from a COVID-19 confirmed case without recommended personal protective equipment (PPE) or with a possible breach of PPE;
- a contact in an aircraft sitting within two seats (in any direction) of a COVID-19 confirmed case, travel companions or persons providing care, and crew members serving in the section of the aircraft where the index case was seated; passengers seated in the entire section or all passengers on the aircraft were considered close contacts of a confirmed case when severity of symptoms or movement of the case indicated more extensive exposure.
Confirmed cases were defined as subjects with positive laboratory confirmation via RT-PCR assays for virus causing of SARS-CoV-2 infection, irrespective of clinical signs and symptoms.

A cluster of contacts was defined as the set of contacts identified by one positive index case. Contact tracing activities were carried out through standardized epidemiological investigations of positive cases (or of their parents/relatives) to determine the history of individuals' exposure. The exposure period was initially defined as the time interval ranging from 14 days before to 14 days after the symptom onset of an index case. Following guidelines from WHO [2], after March 20 the time interval was changed from 2 days before to 14 days after the symptom onset of the index case.

Confirmation of cases was obtained with nasal swabs (UTM viral transport ®, Copan Italia S.p.a) with at least two real-time RT-PCR assays targeting different genes (E and RdRp) of SARS-CoV-2 [3,4]. In addition, a novel quantitative real-time RT-PCR targeting an additional SARS-CoV-2 gene (M) was developed (details provided upon request). From February 21 to February 25, all contacts were tested for the presence of viral genome on their nasopharyngeal trait. From February 26 onward, testing was applied only to symptomatic contacts due to limited availability of testing capacity. From March 20, positivity of nasal swabs was also ascertained from a test that sought a single gene. Individuals with inconclusive tests were swabbed again and re-tested to resolve the diagnosis.

On April 16, the Lombardy Region initiated a large-scale serological screening to evaluate the prevalence of SARS-CoV-2 infection among all remaining untested contacts. Serological screening included both symptomatic and asymptomatic case contacts identified through contact tracing activities without history of a swab for SARS-CoV-2. A relatively small number of serological tests were administered to individuals already tested by RT-PCR (n=627 in our sample). The test used to detect SARS-CoV-2 IgG antibodies is the LIAISON® SARS-CoV-2 test [5,6]. The LIAISON® SARS-CoV-2 test employs magnetic beads coated with S1 & S2 antigens. The antigens used in the tests are expressed in human cells to achieve proper folding, oligomer formation, and glycosylation, providing material similar to the native spikes. This strategy ensures that the antigen-antibody complex forms with the required specificity. The S1 and S2 proteins are both targets to neutralizing antibodies. The test provides the detection of IgG antibodies against S1/S2 antigens of SARS-CoV-2 and the detection of neutralizing antibodies with >97% specificity and >94% sensitivity at 15 days from diagnosis [5,6] and validated against Plaque Reduction Neutralization Test (PRNT) [5]. A negative result (<12 AU/mL) indicates the absence or a very low level of IgG antibodies directed against the virus; this occurs in the absence of infection or during the incubation period or in the early stages of the disease. An inconclusive result (12-15 AU/mL) can be interpreted as both a false positive or a false negative and suggests repeating the exam after a week. A positive result (>15 AU/mL) indicates the presence of IgG antibodies and therefore a past infection with SARS-CoV-2.

## Statistical analysis

Contact tracing data combined with test results and outcomes of close contacts associated with each index case were used to estimate the SARS-CoV-2 infection fatality rate (IFR). The IFR was estimated for male and female individuals separately and for five age groups (0-19 years, 20-49 years, 50-59 years, 60-69 years, 70-79 years, 80+ years) in two distinct epidemic phases (before and after March 16, 2020) and for patients affected by cardiovascular comorbidities (including hypertension) and patients with no comorbidities. The IFR was computed as the proportion of deaths among the total number of infected individuals for each considered strata. Exact binomial test was used to estimate confidence intervals.

To investigate IFR risk factors, we used a generalized linear model (GLM with logit link), using the outcome of positive close contacts as a binary response variable (i.e. death vs. survival) and considering the following covariates:
- the age group of the contact; as no deaths were found below 50 years of age, four age groups were considered in this case: 0-59 years, 60-69 years, 70-79 years, 80+ years.
- the sex of the contact;
- a categorical variable defining whether: 1) the contact was presenting at least one cardiovascular comorbidity, including hypertension; 2) the contact was not presenting cardiovascular comorbidities but was presenting at least one of the following comorbidities: respiratory, oncological, metabolic (including diabetes); 3) no comorbidities were known for the contact;
- a binary variable defining whether the date of identification of the index case associated to the contact was prior or posterior to March 16, 2020; this date represents the median date of confirmation among index cases in the selected clusters (i.e. those with complete testing).

Regression models including interactions between covariates or considering the number of comorbidities affecting a contact were ruled out when compared to model described above on the basis of likelihood ratio tests. Regression models using a numeric variable for the number of comorbidities or a binary variable for the presence of each type of comorbidity instead of the above described categorical variable were ruled out on the basis of the Akaike information criterion.

Risk ratios of death after infection were computed given the covariates, using the generalized linear model described above. Resulting means were compared by Tukey post-hoc test. The statistical analysis was performed using R (version 3.6).

Finally, in the analyzed sample (Table S1), 41.9% (137/327) contacts tested by both RT-PCR and serology resulted negative to RT-PCR and positive to serology. This result may be due to a variety of factors including false positive results from RT-PCR, false negative results from serology, late PCR testing of SARS-CoV-2 infections, transmission occurred outside the analyzed clusters. Although our data are not appropriate to evaluate the accuracy of PCR and serological tests, we performed a sensitivity analysis to explore to explore to what extent false negative PCR results can affect our estimates on the IFR. To do this, we computed the IFR resulting form 10,000 simulations where, a random sample of 41.9% contacts who were RT-PCR negative contacts and were not tested with serology (namely 732) is assumed to be positive to SARS-CoV-2. In this case the IFR results 2.04% (95%CI 1.57-2.60%) instead of the 2.2% (95%CI 1.69-2.81%),obtained in our baseline analysis.

## Supplementary tables and figures

|  | SARS-CoV-2 positive | Deaths | Relative risk |
|---|---|---|---|
| **Sex** |  |  |  |
| Females | 1604 | 29 | reference |
| Males | 1220 | 33 | 1.81 (95%CI 1.03-3.16) |
| **Age** |  |  |  |
| 80+ | 158 | 29 | reference |
| 0-59y | 1837 | 3 | 0.03 (95%CI 0.01-0.1) |
| 60-69y | 494 | 7 | 0.14 (95%CI 0.05-0.32) |
| 70-79y | 335 | 23 | 0.5 (95%CI 0.27-0.89) |
| **Comorbidity** |  |  |  |
| None | 113 | 1 | reference |
| Cardiovascular | 316 | 51 | 5.64 (95%CI 1.17-55.27) |
| Other comorbity | 60 | 1 | 0.93 (95%CI 0.04-20.55) |
| Unknown | 2335 | 9 | 0.36 (95%CI 0.06-6.42) |
| **Epidemic period** |  |  |  |
| Before March 18,2020 | 1423 | 42 | reference |
| After March 18,2020 | 1358 | 18 | 0.38 (95%CI 0.2-0.69) |
| Unknown | 43 | 2 | 3.09 (95%CI 0.43-10.94) |

**Table S1.** Estimated relative risks of death after SARS-CoV-2 infection as obtained by a generalized linear mixed-effects model where the fatal outcome is used as the response variable and sex, age group, comorbidities, and epidemic period are considered as regressors.

| RT-PCR | Serological assay (IgG) | Total |
|---|---|---|
| Performed | Not performed | 1,364 |
| Positive | - | 632 |
| Not performed | Performed | 3,493 |
| - | Positive | 1,755 |
| Performed | Performed | 627 |
| Positive | Negative | 5 |
| Negative | Positive | 137 |
| Positive | Positive | 295 |

**Table S2.** Sample description. In Lombardy, from February 26 onward, only symptomatic contacts of COVID-19 confirmed cases were tested with RT-PCR. Out of 1,991 RT-PCR tests, 1,947 were conducted after February 26.

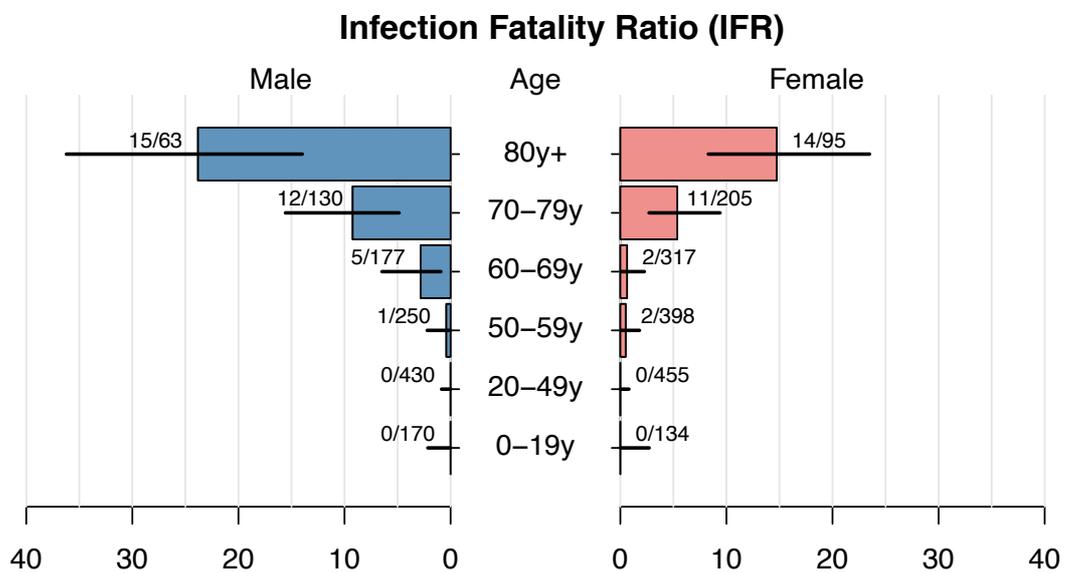

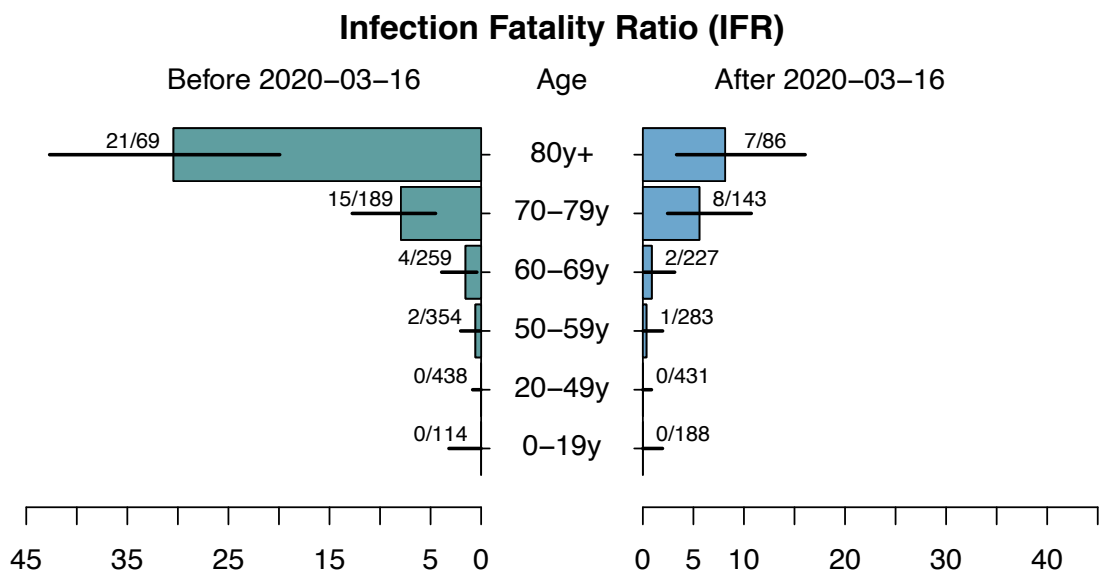

**Figure S1.** Age-specific estimates (mean) of IFR in male and female subjects (top) and during two epidemic periods (bottom). Horizontal lines represent 95% confidence intervals computed by exact binomial tests.

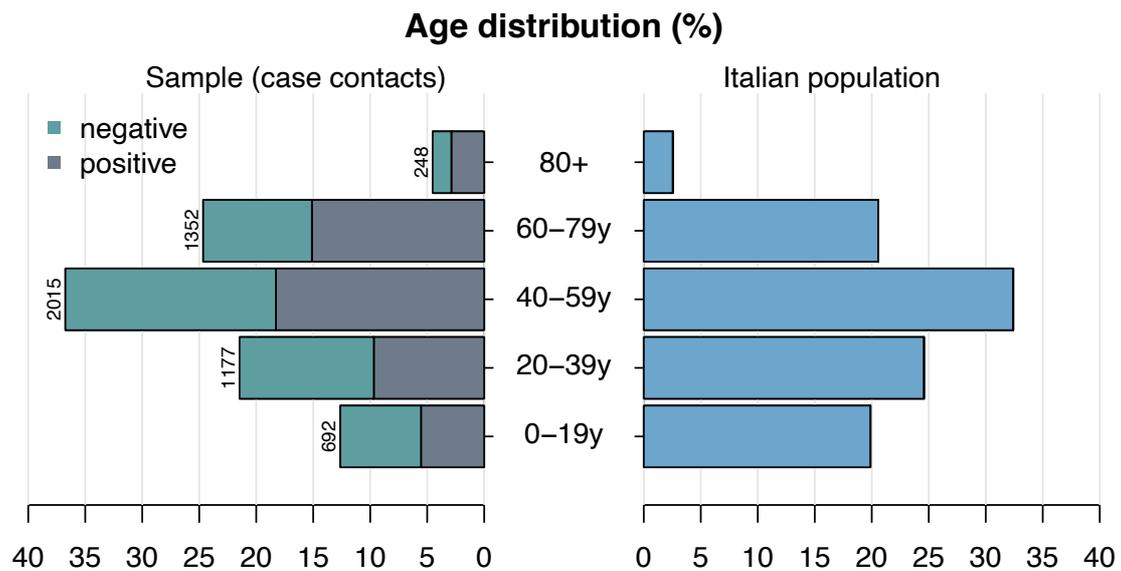

**Figure S2.** Comparison of the age distribution of analyzed close contacts (darker bars represent SARS-CoV-2 positive individuals) with the age distribution of the Lombardy population in 2019, as reported by the Italian National Institute of Statistics [7]